\documentclass[prb,aps,preprint]{revtex4}

\usepackage{graphicx}

\begin{document}

\title{Electronic structure of wurtzite quantum dots with cylindrical symmetry}

\author{L. C. Lew Yan Voon}
\email{lok.lewyanvoon@wright.edu}
\affiliation{Department of Physics, Wright State University, 3640 Colonel Glenn Highway, Dayton, Ohio 45435, USA}

\author{C. Galeriu}
\affiliation{Department of Physics, Worcester Polytechnic Institute, 100 Institute Road, Worcester, Massachusetts 01609, USA}

\author{B. Lassen}
\affiliation{Mads Clausen Institute, University of Southern Denmark, Grundtvigs All\'{e} 150, DK-6400 S\o nderborg, Denmark}

\author{M. Willatzen}
\affiliation{Mads Clausen Institute, University of Southern Denmark, Grundtvigs All\'{e} 150, DK-6400 S\o nderborg, Denmark}

\author{R. Melnik}
\affiliation{Mads Clausen Institute, University of Southern Denmark, Grundtvigs All\'{e} 150, DK-6400 S\o nderborg, Denmark}

\date{\today}

\begin{abstract}
This paper presents a six-band ${\bf k \cdot p}$ theory for wurtzite semiconductor nanostructures with 
cylindrical symmetry. Our work  extends the formulation of Vahala and Sercel 
[{\it Physical Review Letters} {\bf 65}, 239 (1990)] to the Rashba-Sheka-Pikus Hamiltonian 
for wurtzite semiconductors, without the need for the axial approximation.
Results comparing our formulation for studying the electronic structure of wurzite quantum dots 
with the conventional formulation are given.
\end{abstract}


\maketitle

\section{Introduction}

The III-V nitride semiconductors, with wurtzite (WZ) crystal structure, 
have received a great deal of attention in recent years.
In the 1950's there were numerous optical studies of bulk WZ semiconductors
(see Ref.~\onlinecite{LCLYV96} for a recent review). 
However, the detailed band structure of these materials was not studied until the
discovery (in 1993) of blue light emission of WZ GaN on sapphire. \cite{Nakamura00}
Currently the interest in WZ materials has shifted to nanostructures such as
CdSe quantum rods, \cite{Hu01,Li01,Katz02}
ZnS nanowires, \cite{Wang02}
ZnO nanorods \cite{Tian03}
and nanowires, \cite{Huang01}
AlN nanorods \cite{Liu03} and
GaN nanowires. \cite{Johnson02}
On the theory side,
the ${\bf k\cdot p}$ theory of WZ bulk materials was developed by 
Rashba and Pikus \cite{Rashba59,Bir75} and 
later applied to heterostructures by a number of authors.~\cite{Jeon96,Chuang96}
Mireles and Ulloa \cite{Mireles99} have applied the theory to heterostructures
using the envelope function formalism of Burt \cite{Burt92} and Foreman. \cite{Foreman97}
As far as we are aware, the model has only been applied to quantum wells \cite{Chuang97}
and pyramidal quantum dots. \cite{Andreev00}
In 1990, Sercel and Vahala (SV) presented a new formulation of the
multiband envelope function theory and applied it to spherical quantum dots and
cylindrical quantum wires of zincblende (ZB) materials.~\cite{Vahala90,Sercel90b}
For ZB, the SV formulation
was only possible provided the axial approximation 
($\gamma_2=\gamma_3$) was made in the Hamiltonian. 
The theory, within the Luttinger-Kohn (LK) framework, has subsequently been applied to 
model 
quantum rods,~\cite{Katz02}
quantum rings,~\cite{Planelles02a} and
quantum dots.~\cite{Pedersen96}

The current interest in WZ nanowires \cite{Duan00} motivates the need for a model similar
to the one introduced by Sercel and Vahala for ZB, for increasing the physical understanding of the 
${\bf k\cdot p}$ theory and for improving the efficiency of numerical computations for device applications.
In this Letter, we propose a formulation of the 
Rashba-Sheka-Pikus Hamiltonian with cylindrical symmetry.
We have reformulated the ${\bf k\cdot p}$ Hamiltonian in terms of the Sercel-Vahala
(SV) representation for problems with axial symmetry.
The SV representation is useful because it reduces the 3D problem to a 2D one
when cylindrical polar coordinates are used.
Contrary to the work of Sercel and Vahala, where the axial approximation had to be introduced for ZB materials,
we show that the formulation is exact for WZ materials, i.e., 
{\em no axial approximation} was needed. 
In addition to the fundamental interest in the SV representation of the WZ Hamiltonian,
the latter also helps make for more efficient computation for problems with
axial symmetry. This includes free-standing and embedded cylindrical nanowires,
modulated nanowires,
quantum rods, spheroidal and spherical quantum dots.

\section{Hamiltonian}

The six-band Hamiltonian for a WZ semiconductor heterostructure, 
in the $|X\uparrow \rangle, |Y\uparrow \rangle, |Z\uparrow \rangle, 
|X\downarrow\rangle, |Y\downarrow\rangle, |Z\downarrow\rangle$ basis states of the Kane model is \cite{Chuang96,Bir75,Mireles99,Fonoberov03}
\begin{equation}
\widehat{H} = \frac{\hbar^2}{2 m_0} \pmatrix{H_k & 0 \cr 0 & H_k}  + H_{\Delta},
\label{H3D}
\end{equation}
where\cite{Mireles99}
\begin{widetext}
\begin{equation}
H_k = \pmatrix{
\widehat{k}_{x}L_1\widehat{k}_{x}+\widehat{k}_{y}M_1\widehat{k}_{y}+\widehat{k}_{z}M_2\widehat{k}_{z} 
& \widehat{k}_{x}N_1\widehat{k}_{y}+\widehat{k}_{y}N_1'\widehat{k}_{x}
& \widehat{k}_{x}N_2\widehat{k}_{z}+\widehat{k}_{z}N_2'\widehat{k}_{x} \cr
\widehat{k}_{y}N_1\widehat{k}_{x}+\widehat{k}_{x}N_1'\widehat{k}_{y}
& \widehat{k}_{x}M_1\widehat{k}_{x}+\widehat{k}_{y}L_1\widehat{k}_{y}+\widehat{k}_{z}M_2\widehat{k}_{z}
& \widehat{k}_{y}N_2\widehat{k}_{z}+\widehat{k}_{z}N_2'\widehat{k}_{y} \cr
\widehat{k}_{z}N_2\widehat{k}_{x}+\widehat{k}_{x}N_2'\widehat{k}_{z} 
& \widehat{k}_{z}N_2\widehat{k}_{y}+\widehat{k}_{y}N_2'\widehat{k}_{z}
& \widehat{k}_{x}M_3\widehat{k}_{x}+\widehat{k}_{y}M_3\widehat{k}_{y}+\widehat{k}_{z}L_2\widehat{k}_{z} 
\label{Hk}
},
\end{equation}
\begin{equation}
H_{\Delta} = \bordermatrix{
& |X\uparrow \rangle & |Y\uparrow \rangle & |Z\uparrow \rangle 
& |X\downarrow\rangle & |Y\downarrow\rangle & |Z\downarrow\rangle \cr
& \Delta_1 & -i \Delta_2 & 0 & 0 & 0 & \Delta_3 \cr
& i \Delta_2 & \Delta_1 & 0 & 0 & 0 & -i \Delta_3 \cr
& 0 & 0 & 0 & -\Delta_3 & i \Delta_3 & 0 \cr
& 0 & 0 & -\Delta_3 & \Delta_1 & i \Delta_2 & 0 \cr
& 0 & 0 & -i \Delta_3 & -i \Delta_2 & \Delta_1 & 0  \cr
& \Delta_3  & i \Delta_3 & 0 & 0 & 0 & 0 
\label{HDelta}
}.
\end{equation}
\end{widetext}

The $L_i$'s, $M_i$'s, $N_i$'s and $N_i'$'s are the Burt-Foreman valence-band $k\cdot p$ parameters
introduced by Mireles and Ulloa\cite{Mireles99}; the $\Delta_i$'s are spin-orbit 
parameters.\cite{Mireles99}
Next, one can use the $|u_1\rangle,|u_2\rangle,|u_3\rangle,|u_4\rangle,|u_5\rangle,|u_6\rangle$ basis states defined 
in~Ref.\protect\onlinecite{Chuang96}.
The advantage of using this basis is that now the zone-center Bloch functions are described 
by an angular momentum {\bf $J$}.
The envelope part of the total wave function 
behaves like the spherical harmonics, with an angular momentum {\bf $L$}.
The periodic Bloch space and the slowly varying envelope space are coupled by the ${\bf k\cdot p}$ interaction.
For problems with cylindrical symmetry, the projection of the total angular momentum $F_z = L_z + J_z$
is a good quantum number. 

Next, we express the ${\bf k\cdot p}$ Hamiltonian in terms of cylindrical polar coordinates $\rho, \phi, z$. 
The total wave function is then written as
\begin{eqnarray}
&&\psi({\bf r}) 
= \sum_{i} f_i({\bf r})|u_i\rangle
= \sum_{i} g_{i}(\rho,z) e^{i L_{z i} \phi}|u_i\rangle \nonumber \\
&&= \sum_{i} g_{i}(\rho,z) e^{i(F_z-J_{z i})\phi}|u_i\rangle
= \sum_{i} g_{i}(\rho,z)|u_i'\rangle.
\end{eqnarray}
There is a double degeneracy with respect to the sign of $F_z$ 
due to inversion and time-reversal symmetry. The new matrix elements are given by
\begin{eqnarray}
&&\langle f_i({\bf r})|\widehat{H}|f_j({\bf r})\rangle \nonumber \\
&&= \langle g_i(\rho,z)|e^{-i(F_z-J_{z i})\phi}\widehat{H}e^{i(F_z'-J_{z j})\phi}|g_j(\rho,z)\rangle \nonumber \\
&&= \langle g_i(\rho,z)|\widehat{H'}|g_j(\rho,z)\rangle.
\end{eqnarray}
These matrix elements are zero unless $F_z = F_z'$.

It is found that all $\phi$ dependence goes away following the SV transformation,
without making an axial approximation as for ZB. 
The validity of this result is due to the axial symmetry already found to be true
for the bulk dispersion relation.~\cite{Chuang96}
This can also be seen from a group theoretic point of view by noting that
the group of the wave vector is the same for all wave vectors in the plane
perpendicular to the $c$ axis.
Finally, the WZ Hamiltonian in cylindrical coordinates is 
\begin{widetext}
\begin{equation}
\widehat{H'} = \bordermatrix{
& |u_1'\rangle & |u_2'\rangle & |u_3'\rangle & |u_4'\rangle & |u_5'\rangle & |u_6'\rangle \cr
& { S_{11} \atop + \Delta_1 + \Delta_2 }
& S_{12} 
& S_{13} 
& 0 & 0 & 0 \cr
&&&&&& \cr
& S_{21} 
& { S_{22} \atop + \Delta_1 - \Delta_2 }
& S_{23} 
& 0 & 0 & \sqrt{2} \Delta_3 \cr
&&&&&& \cr
& S_{31} 
& S_{32} 
& S_{33} 
& 0 & \sqrt{2} \Delta_3 & 0 \cr
&&&&&& \cr
& 0 & 0 & 0 
& { S_{44} \atop + \Delta_1 + \Delta_2 }
&S_{45} 
& S_{46} \cr
&&&&&& \cr
& 0 & 0 & \sqrt{2} \Delta_3 
& S_{54} 
& { S_{55} \atop + \Delta_1 - \Delta_2 }
& S_{56} \cr
&&&&&& \cr
& 0 & \sqrt{2} \Delta_3 & 0 
& S_{64} 
& S_{65} 
& S_{66} \cr
},
\label{H2D}
\end{equation}
\end{widetext}
where
\begin{eqnarray}
S_{11} &=& -\frac{\hbar^2}{2m_0}\frac{1}{2}\Bigg\{
\frac{\partial}{\partial \rho}\left((L_1+M_1)\frac{\partial}{\partial \rho}\right)
+ \frac{(L_1+M_1)}{\rho}\frac{\partial}{\partial \rho} \nonumber \\
&+& 2\frac{\partial}{\partial z}M_2\frac{\partial}{\partial z}
- \frac{(F_z-J_1)}{\rho}\frac{\partial (N_1-N_1')}{\partial \rho} - \frac{(F_z-J_1)^2}{\rho^2}(L_1+M_1)
\Bigg\}, \nonumber\\
S_{22} &=& -\frac{\hbar^2}{2m_0}\frac{1}{2}\Bigg\{
  \frac{\partial}{\partial \rho}\left((L_1+M_1)\frac{\partial}{\partial \rho}\right)
+ \frac{(L_1+M_1)}{\rho}\frac{\partial}{\partial \rho} \nonumber \\
&+& 2\frac{\partial}{\partial z}M_2\frac{\partial}{\partial z}
+ \frac{(F_z-J_2)}{\rho}\frac{\partial (N_1-N_1')}{\partial \rho} - \frac{(F_z-J_2)^2}{\rho^2}(L_1+M_1)
\Bigg\}, \nonumber\\
S_{33} &=& -\frac{\hbar^2}{2m_0}\Bigg\{
  \frac{\partial}{\partial \rho}\left(M_3\frac{\partial}{\partial \rho}\right)
+ \frac{M_3}{\rho}\frac{\partial}{\partial \rho} + \frac{\partial}{\partial z}L_2\frac{\partial}{\partial z}
- \frac{(F_z-J_3)^2}{\rho^2}M_3
\Bigg\}, \nonumber\\
S_{12} &=& \frac{\hbar^2}{2m_0}\frac{1}{2}\Bigg\{
\frac{\partial}{\partial\rho}\left((L_1-M_1)\frac{\partial}{\partial\rho}\right) 
+ (F_z-J_2)\frac{\partial}{\partial\rho}\left(\frac{(L_1-M_1)}{\rho}\cdot\right) \nonumber \\
&+& (F_z-J_2-1)\frac{(L_1-M_1)}{\rho}\frac{\partial}{\partial\rho} + \frac{(F_z-J_2)(F_z-J_2-1)}{\rho^2}(L_1-M_1)
\Bigg\}, \nonumber\\
S_{21} &=& \frac{\hbar^2}{2m_0}\frac{1}{2}\Bigg\{
\frac{\partial}{\partial\rho}\left((L_1-M_1)\frac{\partial}{\partial\rho}\right)
- (F_z-J_1)\frac{\partial}{\partial\rho}\left(\frac{(L_1-M_1)}{\rho}\cdot\right) \nonumber \\
&-& (F_z-J_1+1)\frac{(L_1-M_1)}{\rho}\frac{\partial}{\partial\rho} + \frac{(F_z-J_1)(F_z-J_1+1)}{\rho^2}(L_1-M_1)
\Bigg\}, \nonumber\\
S_{13} &=& \frac{\hbar^2}{2m_0}\frac{1}{\sqrt{2}}\Bigg\{
  \frac{\partial}{\partial \rho}\left(N_2\frac{\partial}{\partial z}\right)
+ \frac{\partial}{\partial z}\left(N_2'\frac{\partial}{\partial \rho}\right) 
+ (F_z-J_3)\left[ \frac{\partial}{\partial z}\left(\frac{N_2'}{\rho}\cdot\right)
 + \frac{N_2}{\rho}\frac{\partial}{\partial z}\right]
\Bigg\}, \nonumber\\
S_{31} &=& \frac{\hbar^2}{2m_0}\frac{1}{\sqrt{2}}\Bigg\{
  \frac{\partial}{\partial\rho}\left(N_2'\frac{\partial}{\partial z}\right)
+ \frac{\partial}{\partial z}\left(N_2\frac{\partial}{\partial \rho}\right) 
- (F_z-J_1)\left[ \frac{\partial}{\partial z}\left(\frac{N_2}{\rho}\cdot\right)
 + \frac{N_2'}{\rho}\frac{\partial}{\partial z}\right]
\Bigg\}, \\
S_{23} &=& -\frac{\hbar^2}{2m_0}\frac{1}{\sqrt{2}}\Bigg\{
  \frac{\partial}{\partial \rho}\left(N_2\frac{\partial}{\partial z}\right)
+ \frac{\partial}{\partial z}\left(N_2'\frac{\partial}{\partial \rho}\right) 
- (F_z-J_3)\left[ \frac{\partial}{\partial z}\left(\frac{N_2'}{\rho}\cdot\right)
 + \frac{N_2}{\rho}\frac{\partial}{\partial z}\right]
\Bigg\}, \nonumber\\
S_{32} &=& -\frac{\hbar^2}{2m_0}\frac{1}{\sqrt{2}}\Bigg\{
  \frac{\partial}{\partial \rho}\left(N_2'\frac{\partial}{\partial z}\right)
+ \frac{\partial}{\partial z}\left(N_2\frac{\partial}{\partial \rho}\right) 
+ (F_z-J_2)\left[ \frac{\partial}{\partial z}\left(\frac{N_2}{\rho}\cdot\right)
 + \frac{N_2'}{\rho}\frac{\partial}{\partial z}\right]
\Bigg\}, \nonumber\\
S_{44} &=& -\frac{\hbar^2}{2m_0}\frac{1}{2}\Bigg\{
  \frac{\partial}{\partial \rho}\left((L_1+M_1)\frac{\partial}{\partial \rho}\right)
+ \frac{(L_1+M_1)}{\rho}\frac{\partial}{\partial \rho} \nonumber \\
&+& 2\frac{\partial}{\partial z}M_2\frac{\partial}{\partial z}
+ \frac{(F_z-J_4)}{\rho}\frac{\partial (N_1-N_1')}{\partial \rho} - \frac{(F_z-J_4)^2}{\rho^2}(L_1+M_1)
\Bigg\}, \nonumber\\
S_{55} &=& -\frac{\hbar^2}{2m_0}\frac{1}{2}\Bigg\{
  \frac{\partial}{\partial \rho}\left((L_1+M_1)\frac{\partial}{\partial \rho}\right)
+ \frac{(L_1+M_1)}{\rho}\frac{\partial}{\partial \rho} \nonumber \\
&+& 2\frac{\partial}{\partial z}M_2\frac{\partial}{\partial z}
- \frac{(F_z-J_5)}{\rho}\frac{\partial (N_1-N_1')}{\partial \rho} - \frac{(F_z-J_5)^2}{\rho^2}(L_1+M_1)
\Bigg\}, \nonumber\\
S_{66} &=& -\frac{\hbar^2}{2m_0}\Bigg\{
  \frac{\partial}{\partial \rho}\left(M_3\frac{\partial}{\partial \rho}\right)
+ \frac{M_3}{\rho}\frac{\partial}{\partial \rho} + \frac{\partial}{\partial z}L_2\frac{\partial}{\partial z}
- \frac{(F_z-J_6)^2}{\rho^2}M_3
\Bigg\}, \nonumber\\
S_{45} &=& \frac{\hbar^2}{2m_0}\frac{1}{2}\Bigg\{
\frac{\partial}{\partial\rho}\left((L_1-M_1)\frac{\partial}{\partial\rho}\right) 
- (F_z-J_5)\frac{\partial}{\partial\rho}\left(\frac{(L_1-M_1)}{\rho}\cdot\right) \nonumber \\
&-& (F_z-J_5+1)\frac{(L_1-M_1)}{\rho}\frac{\partial}{\partial\rho} + \frac{(F_z-J_5)(F_z-J_5+1)}{\rho^2}(L_1-M_1)
\Bigg\}, \nonumber\\
S_{54} &=&  \frac{\hbar^2}{2m_0}\frac{1}{2}\Bigg\{
\frac{\partial}{\partial\rho}\left((L_1-M_1)\frac{\partial}{\partial\rho}\right) + (F_z-J_4)\frac{\partial}{\partial\rho}\left(\frac{(L_1-M_1)}{\rho}\cdot\right) \nonumber \\
&+& (F_z-J_4-1)\frac{(L_1-M_1)}{\rho}\frac{\partial}{\partial\rho} + \frac{(F_z-J_4)(F_z-J_4-1)}{\rho^2}(L_1-M_1)
\Bigg\}, \nonumber\\
S_{46} &=& -\frac{\hbar^2}{2m_0}\frac{1}{\sqrt{2}}\Bigg\{
  \frac{\partial}{\partial \rho}\left(N_2\frac{\partial}{\partial z}\right)
+ \frac{\partial}{\partial z}\left(N_2'\frac{\partial}{\partial \rho}\right) - (F_z-J_6)\left[ \frac{\partial}{\partial z}\left(\frac{N_2'\cdot}{\rho}\right)
 + \frac{N_2}{\rho}\frac{\partial}{\partial z}\right]
\Bigg\}, \nonumber\\
S_{56} &=& \frac{\hbar^2}{2m_0}\frac{1}{\sqrt{2}}\Bigg\{
  \frac{\partial}{\partial \rho}\left(N_2\frac{\partial}{\partial z}\right)
+ \frac{\partial}{\partial z}\left(N_2'\frac{\partial}{\partial \rho}\right) + (F_z-J_6)\left[ \frac{\partial}{\partial z}\left(\frac{N_2'\cdot}{\rho}\right)
 + \frac{N_2}{\rho}\frac{\partial}{\partial z}\right]
\Bigg\}, \nonumber\\
S_{64} &=& -\frac{\hbar^2}{2m_0}\frac{1}{\sqrt{2}}\Bigg\{
  \frac{\partial}{\partial \rho}\left(N_2'\frac{\partial}{\partial z}\right)
+ \frac{\partial}{\partial z}\left(N_2\frac{\partial}{\partial \rho}\right) + (F_z-J_4)\left[ \frac{\partial}{\partial z}\left(\frac{N_2\cdot}{\rho}\right)
 + \frac{N_2'}{\rho}\frac{\partial}{\partial z}\right]
\Bigg\}, \nonumber\\
S_{65} &=& \frac{\hbar^2}{2m_0}\frac{1}{\sqrt{2}}\Bigg\{
  \frac{\partial}{\partial \rho}\left(N_2'\frac{\partial}{\partial z}\right)
+ \frac{\partial}{\partial z}\left(N_2\frac{\partial}{\partial \rho}\right) - (F_z-J_5)\left[ \frac{\partial}{\partial z}\left(\frac{N_2\cdot}{\rho}\right)
 + \frac{N_2'}{\rho}\frac{\partial}{\partial z}\right]
\Bigg\}, \nonumber
\end{eqnarray}
where $J_n \equiv J_{z n}$.
Note that in general $S_{ij} \neq S_{ji}^\dagger$. That the matrix representation is
nonhermitian is due to the chosen representation and to the fact that the Hilbert space
is now divided into $F_z$ subspaces. An analogous result was obtained previously
for the SV representation of zincblende quantum rings.\cite{Planelles02a}

\section{Numerical Results and Discussion}

\begin{figure}
\includegraphics{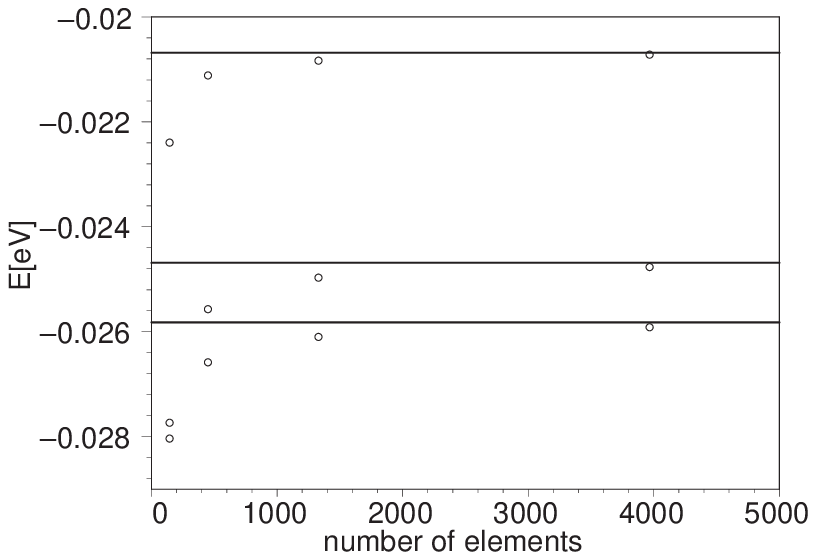}
\caption{\label{fig:fig1} The energies of the top three valence sublevels 
of a WZ GaN cylindrical quantum dot, calculated with the 3D program (circles).
The solid lines are the converged values obtained with the 2D program.}
\end{figure}

A free standing WZ GaN cylindrical quantum dot, 
with a radius of 50 \AA\ and a height of 100 \AA, 
with the axis along the crystallographic $c$-axis, has been studied using
both the 3D and 2D methods. 
The parameters for GaN have been taken from the literature.\cite{Fonoberov03}
The 3D Hamiltonian (\ref{H3D}) and the 2D Hamiltonian (\ref{H2D}) have been implemented 
using FEMLAB; this is a software using the finite element method.
Different meshes have been generated, and the convergence of the eigenvalues has been studied.
Due to the reduced dimensionality, the eigenvalue problem in 2D can be solved
much faster, in minutes instead of hours, and with less memory requirements than
the eigenvalue problem in 3D. After only one mesh refinement 
(2040 elements) 
the 2D eigenvalues have converged. However, even after three mesh refinements 
(3967 elements) the 3D
eigenvalues are not fully converged, as seen from Fig. \ref{fig:fig1}.

\begin{figure}
\includegraphics{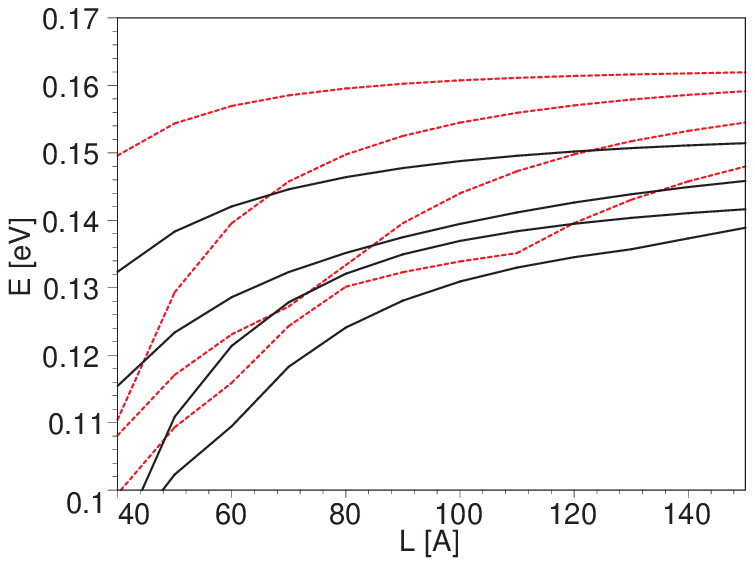}
\caption{\label{fig:fig2} The energies of the lowest states as a function of the CdSe quantum dot aspect ratio. 
The linear term is not included in the Hamiltonian.
Shown are the first four states with $F_z=1/2$ (solid lines) and with $F_z=3/2$ (dotted lines).}
\end{figure}

\begin{figure}
\includegraphics{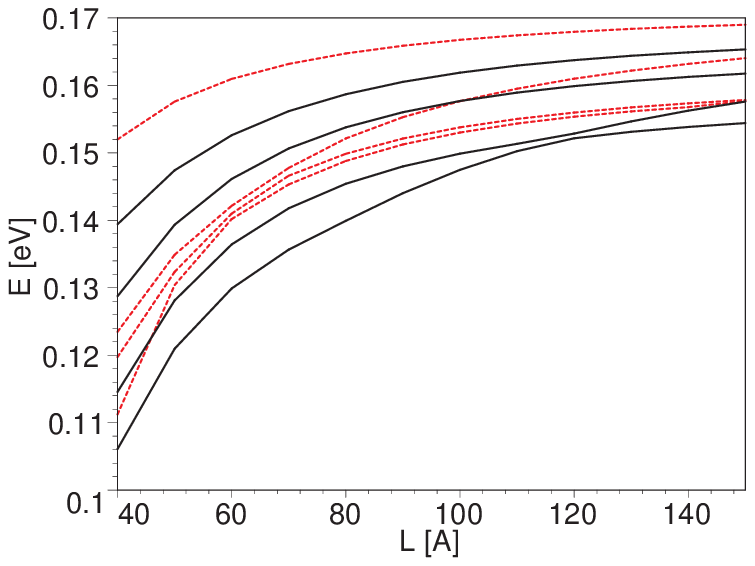}
\caption{\label{fig:fig3} The energies of the lowest states as a function of the CdSe quantum dot aspect ratio. 
The linear term is included in the Hamiltonian.
Shown are the first four states with $F_z=1/2$ (solid lines) and with $F_z=3/2$ (dotted lines).}
\end{figure}

A free standing WZ CdSe cylindrical quantum dot, 
with a radius of 50 \AA\ and a height of 40--150 \AA, 
has also been investigated.
The parameters for CdSe have been taken from the literature.\cite{Jeon96,Li02}
By varying the height we can study shape effects on semiconductor nanocrystals.
From Figures \ref{fig:fig2}-\ref{fig:fig3}
we notice that there
are crossings between states with $F_z=1/2$ (solid lines) and $F_z=3/2$ (dotted lines).
The inclusion of the linear term in the Hamiltonian dramatically changes the 
energy band structure, and results in anticrossings between energy bands \cite{Xia99a}.

In CdSe nanocrystals with a radius of 15 \AA\ and a gelcap shape a crossing
between the lowest two valence sublevels
appears at an aspect ratio of 1.25. \cite{Hu01}
There are also some experimental studies of the transition
from 3D to 2D confinement in CdSe quantum rods.~\cite{Hu01,Yu03}
Theoretical work on the electronic properties of the transition reported so far
include empirical pseudopotential calculations on a rather artificial shape,~\cite{Hu01}
and $\bf k\cdot p$ calculations for cylindrical ZB structures~\cite{Katz02,LYV04NL}
and for spheroidal WZ structures.~\cite{Li02}
We hope that our formulation for studying the electronic structure of wurzite quantum dots
with cylindrical symmetry will be a very useful tool in this important research area. 

\section{Conclusions}

We have extended the Sercel-Vahala technique to WZ heterostructures with cylindrical symmetry
and shown this to be an exact result.
This is crucial for applications in enabling greatly decreased 
computational resources. 
We have studied GaN and CdSe
cylindrical quantum dots, calculating the effect of the aspect ratio
on energy levels and wavefunctions.  

\begin{acknowledgments}
The work was supported by an NSF CAREER award (NSF Grant No. 9984059, 0454849),
and by a Research Challenge grant from Wright State University and the Ohio
Board of Regents. 
\end{acknowledgments}


\newpage

\end{document}